\documentclass[12pt,fleqn]{article}
\usepackage{amsmath}
\usepackage{amsfonts}
\usepackage{amssymb}
\usepackage{graphics,graphicx,epstopdf}
\usepackage{cite}
\usepackage{latexsym}

\begin{document}

\title{Scalar field localization \\ on deformed extra space}
\author{Sergey G. Rubin \\
National Research Nuclear University "MEPhI", \\ (Moscow Engineering Physics Institute)  \\
sergeirubin@list.ru}

\maketitle

\begin{abstract}
Field localization on 2-dim extra space is considered in the framework of $f(R)$ gravity. It is shown that an interference of local matter energy distribution and a metric of the extra space form a point-like defect - 4-dim brane. The energy-momentum of the brane depends on initial conditions that could lead to the cosmological $\Lambda$ term being arbitrarily small.   
\end{abstract}

\section{Introduction}

It is well known that observed physical parameters of the low energy theory differ in many orders of magnitude. Otherwise, no complex structures would be formed. At the same time a future theory should not contain very big/small physical parameters. The fine-tuning (FT) of observable parameters makes the situation much more challenging. Indeed most probable basis of FT, the multiverse idea implies an existence of huge variety of universes with different properties.  At the first glance the multiverse can not be obtained having fixed set of initial parameters of future theory.
There are a lot of attempts to explain the observable Universe with moderate success \cite{Chacko}, \cite{Das}. One of the most known ways is based on the idea of extra space and more specifically on 4-dim branes embedded in n-dim extra space \cite{Rubakov}, \cite{Polch}.

Physical parameters of a primary theory are not only ones which influence the low energy physics. Initial conditions of a universe formation  play significant role also. Well known idea of the Universe creation from a space-time foam presumes randomly distributed initial conditions. An accidental formation of manifolds with various metrics and topologies may be considered as a source of different universes whose variety is connected with a huge number of stationary metrics of an extra space \cite{Brand}, \cite{RuZin}. 
Even if parameters of a primary Lagrangian are fixed we still have a set of initial conditions provided by the space-time foam.
It could resolve the FT and hierarchy problems. 

The probability of the Universe birth was calculated via numerous approaches with radically 
different results \cite{Vilenkin}. It is not surprising due to the absence of the theory of Quantum Gravity. In this study it will be sufficient to assume  nonzero probability of any metric originated from space-time foam. Some of these metrics evolve classically to stationary states, see e.g. \cite{2006PhRvD..73l4019B}.

The idea of random initial conditions originated from the space-time foam are used in this paper to explain the  smallness of the cosmological constant, the prominent example of the fine tuning. 

The extra spaces with point-like defects \cite{RuRole} underlie this research. It is shown that a  scalar field is localized on the point-like defects of 2-dim metric of extra space while its distribution in ordinary space is assumed to be uniform. In its turn back reaction of the trapped scalar field on the extra space metric appears to be meaningful.  It deepens the "gravitational well"\, as it occurs in case of the Einstein gravity. 

The point-like defects being uniformly distributed in our space form  a brane embedded in $(4+d)$-dim space. Brane solutions with various ways of matter localization have been widely investigated in the literature where brane existence is usually postulated \cite{Oda}, \cite{Navarro}. Another widely used way to form a brane in noncompact extra spaces is to involve scalar fields with a complex potential \cite{Shap}.  
Matter fields are trapped by such branes and deform their shape. This result may be obtained in the simplest version of $f(R)$ gravity, see e.g. \cite{Chinaset}.  Branes embedded in 6-dim space are also the subjects of wide discussion \cite{Kogan}, \cite{Giovannini},\cite{Navarro1}.

In this paper the smallness of the $\Lambda$ term produced by the brane is explained without fine tuning of physical parameters. It means that all physical parameters of a Lagrangian are chosen within the interval $10^{-3}<\xi <10^3$ if it is dimensionless unit, or it lies within the interval  $(10^{-3}m_D)^n<\xi <(10^3 m_D)^n$ if its dimensionality is $[m_D]^n$.

The general basis of present study is $f(R)$ gravity. The interest in $f(R)$ theories is motivated by inflationary scenarios starting with the pioneering work of Starobinsky \cite{Starob}. A number of viable $f(R)$ models in 4-dim space that satisfy the observable constraints are proposed in Refs. \cite{Amendola}, \cite{Starob1}, \cite{Odin}. 

Non-linear Lagrangians depending on the Ricci invariants inevitably appear if one takes into account quantum phenomena. \cite{Grib}, \cite{Birrell}, \cite{Donoghue}. At first sight there are no reasons to set $f(R)=R$ if it does not contradict the observations. However, more thorough analysis indicates serious internal problems. They include negative metric states, unitarity violation,  and instability of flat space. However Simon \cite{Simon} has shown that
these problems do not appear when the theory is treated as an effective field theory. In the
low energy limit the effect of nonlinearity is a small correction to the pure Einstein-Hilbert action and no bad behavior is revealed. This particularly relates to the Kaluza-Klein $f(R)$ gravity where the small parameter - ratio of curvature in our 4-dim space and a curvature of compact extra space - arises naturally.

The following sections are devoted to the mechanism of the point-like defects formation.

\section{Setup}

From here on, it is assumed that a characteristic scale of extra space is small and its geometry has been stabilized shortly after  the Universe creation. The stabilization issue is discussed in   \cite{Green}, \cite{KKR}.

As a common basis, consider a Riemannian manifold with metric
\begin{equation}\label{metric}
ds^2 = \mathfrak{G}_{AB}dZ^A dZ^B = g_{\mu\nu}(x)dx^{\mu}dx^{\nu} + G_{ab}(x,y)dy^a dy^b
\end{equation} 
Here $M$, $M'$ are the manifolds with metrics $g_{mn}(x)$ and $G_{ab}(x,y)$ respectively.  $x$ and $y$ are the coordinates of the subspaces $M$ and $M'$. We will refer to 4-dim space $M$ and $n$-dim compact space $M'$ as a main space and an extra space respectively. Here the metric has the signature (+ - - - ...), the
Greek indices $\mu, \nu =0,1,2,3$ refer to 4-dimensional coordinates). Latin indices run
over $a,b, ... = 4, 5...$.

A time behavior of the metric tensor $G_{ab}(x,y)$ is governed by the classical equations of motion and changes under variations of initial conditions. As was shown in \cite{KKR} the energy dissipation into the main space $M$ leads to an entropy decrease of the manifold $M'$. This explains an emergence of a friction term in the classical equations for the extra metric $G_{ab}(x,y)$. This term stabilizes the extra metric. Finally the inflationary process strongly smooths out space inhomogeneity so that
\begin{equation}\label{uniform}
G_{ab}(x,y)
\xrightarrow{t\rightarrow\infty}G_{ab} (t,y)
\end{equation}
at the modern epoch. Time dependence of the external metric was discussed within the framework of the Kaluza-Klein cosmology and Einstein's gravity \cite{Abbott}. If a gravitational Lagrangian contains terms nonlinear in the Ricci scalar the extra metric $G_{ab}$ could have asymptotically stationary states \cite{2006PhRvD..73l4019B},  \cite{KKR} 
\begin{equation}\label{statio}
G_{ab}(t,y)\rightarrow G_{ab}(y),
\end{equation}
see also \cite{Carroll}, \cite{Nasri} for discussion.

Let us estimate the rate of stabilization of the
extra space. Weak deviations of the geometry from an equilibrium configuration can be interpreted as excited states with the mass $m_{KK}$,(see, for example, \cite{Antoniadis}). Since it is the only scale, the decay probability is expected to satisfy the relationship $\Gamma \sim m_{KK}\sim
1/L_n$, where $L_n$ is the characteristic size
of the n-dim extra space.  According to observations $L_n \leq 10^{-18}$cm so that the lifetime of the excited state is $t_1 \sim L_n \leq 10^{-28}$s. Therefore, the extra space reaches a stationary state long before the onset of the primordial nucleosynthesis but, possibly, after completion of the inflationary stage.

According to \eqref{metric}, \eqref{statio} the Ricci scalar represents a simple sum of the Ricci scalar of the main space and the Ricci scalar of extra space
\begin{equation}
R=R_4 + R_n . 
\end{equation}
In the following natural inequality
\begin{equation}\label{ll}
R_4 \ll R_n
 \end{equation}
is assumed. This suggestion looks natural for the extra space size $L_n < 10^{-18}$ cm as
compared to the Schwarzschild radius $L_n \ll r_g \sim 10^6$cm of stellar mass black hole
where the largest curvature in the modern Universe exists. 

Consider a gravity with higher order derivatives and the action in the form,
\begin{eqnarray}\label{act1}
&& S=\frac{m_D ^{D-2}}{2}\int d^{D}Z \sqrt{|\mathfrak{G}|}\left[f(R)+L _m\right]; \\
&& f(R) = \sum\limits_k {a_k R^k } \nonumber 
\end{eqnarray} 
with arbitrary parameters $a_k,\, k\neq 1$ and $a_1 = 1$. Here $D=n+4$ and $L_m$ is a Lagrangian of matter.

Using inequality \eqref{ll} the Taylor expansion of $f(R)$ in  Eq. \ref{act1} gives
\begin{eqnarray}\label{act2}
&& S= \frac{m_D ^{D-2}}{2}\int d^4 x d^n y \sqrt{|g(x)|} \sqrt{|G(y)|}\left[ f(R_4 + R_n ) +L_m \right]\\
&& \simeq \frac{m_D ^{D-2}}{2}\int d^4 x d^n y\sqrt{|g(x)|} \sqrt{|G(y)|}[ R_4(x) f' (R_n (y) ) + f(R_n (y))+L_m]  \nonumber 
 \end{eqnarray}

In this paper a stationary and uniform distribution of the matter fields is taken into account. Comparison of the second line in expression \eqref{act2} with the Einstein-Hilbert action
\begin{equation}
S_{EH}=\frac{M^2 _{Pl}}{2} \int d^4x  \sqrt{|g(x)|}(R-2\Lambda)
\end{equation}
gives the expression  
\begin{equation}\label{MPl}
M^2 _{Pl}=m_{D} ^{D-2}\int d^n y  \sqrt{|G(y)|}f' (R_n (y) )
\end{equation}
for the Planck mass. According to effective the action \eqref{act2}  the term
\begin{equation}\label{density}
\Lambda \equiv -\frac{m_D ^{D-2}}{2M^2 _{Pl}}\int d^n y \sqrt{|G(y)|} [f(R_n) + L_m (y)]
\end{equation}
represents the cosmological $\Lambda$ term. Both The Planck mass and the $\Lambda$ term depends on a stationary geometry $G_{ab} (y)$.  As shown below physical parameters $\{a_k\}$ are not necessary fine tuned in order to make the $\Lambda$ term small.

\section{Deformed extra space and boundary conditions}

Let us find the 2-dim metric disregarding the influence of scalar field and keeping in mind \eqref{ll}. In this work, the metric formalism is used, which consists of varying the action with respect to $G^{ab}$. Note that r.h.s. of expression \eqref{act2} consists of three terms. The first term is much smaller than the second one according to \eqref{ll}; accordingly, stationary configuration $G_{ab} (y)$ is determined by static classical equations  
 \begin{equation}\label{eqS1}
 \frac{\delta S}{\delta G^{ab} (y)}\simeq \frac{\delta S_{extra}}{\delta G^{ab} (y)}=0, \quad S_{extra} = \frac{m_D ^{D-2}}{2}v_4 \int d^ny\sqrt{|G|  }[ f(R_n)+L_m],
 \end{equation}
where $v_4 =\int d^4x  \sqrt{g(x)}$, or in more explicit form
\begin{equation}\label{eqn}
R_{ab} f' -\frac{1}{2}f(R)G_{ab} 
- \nabla_a\nabla_b f_R + G_{ab} \square f' = \frac{1}{m_D^{D-2}}T_{ab}.
\end{equation}
Here $\square$ stands by the d'Alembert operator
\begin{equation}
\square =\square_n =\frac{1}{\sqrt{|G|}}\partial_a ( {G}^{ab}\sqrt{|G|}\partial_b),\quad a,b=1,2.
\end{equation}
Here, the term proportional to $R_4$ is omitted, see \eqref{ll}.
The smallness of the Lambda-term \eqref{density} is the additional condition that will be proven later.

Evidently, there is continuum set of solutions to  system \eqref{eqn} depending on boundary conditions. Maximally symmetrical extra spaces which are used in great majority of literature represent a small subset of the continuum set. The relationship of the extra metric and boundary conditions is studied below. We will see that such nontrivial metrics lead to interesting results.

The trace of \eqref{eqn} can be written in the form
\begin{equation}\label{tr-n}
f'(R_n ) R_n-\frac{n}{2} f(R_n ) +  (n-1)\square_n  f'(R_n ) = T,
\end{equation}
which will be used below. From here on we will use the units $m_D = 1$. 

In order to perform numerical analysis in the first order approximation let us disregard the matter contribution and choose $n=2$ which strongly facilitates the analysis. Indeed, if an extra space is 2-dimensional, only one equation in system \eqref{eqn} remains independent. Let it be equation \eqref{tr-n}. 

The compact 2-dim manifold is supposed to be parameterized by the two spherical angles $\theta$ and $\phi$ $(0 \leq\theta \leq \pi, 0 \leq \phi \leq  2\pi)$.
The choice of metric
\begin{equation}\label{metric2}
G_{\theta\theta} = -r(\theta)^2;\quad G_{\phi \phi} -r(\theta)^2 \sin^2(\theta).
\end{equation}
leads to the Ricci scalar expressed in terms of the radius $r(\theta)$
\begin{equation}\label{Ricci}
R=\frac{2}{r(\theta)^4\sin(\theta)}(-r'r\cos(\theta)+r^2 \sin(\theta)+r'^2 \sin(\theta)-\sin(\theta)rr''),
\end{equation}
where prime means $d/d\theta$.

If a brane is artificially attached to a point $\theta =0$ the 2-dim metric acquires the form
\begin{equation}
r_0^2(d\theta^2 + \alpha^2 sin^2\theta d\phi^2)
\end{equation}
The effect of the brane causes a deficit angle
$\delta = 2\pi (1-\alpha)$ in a bulk with flat 4-dim metric. The cost is strict connection between parameters of model, see e.g. \cite{Navarro},\cite{Giovannini} and review \cite{Maartens} where pros and cons of this approach are discussed. 

\begin{figure}[t]
\centering
\includegraphics[width=0.7\linewidth]{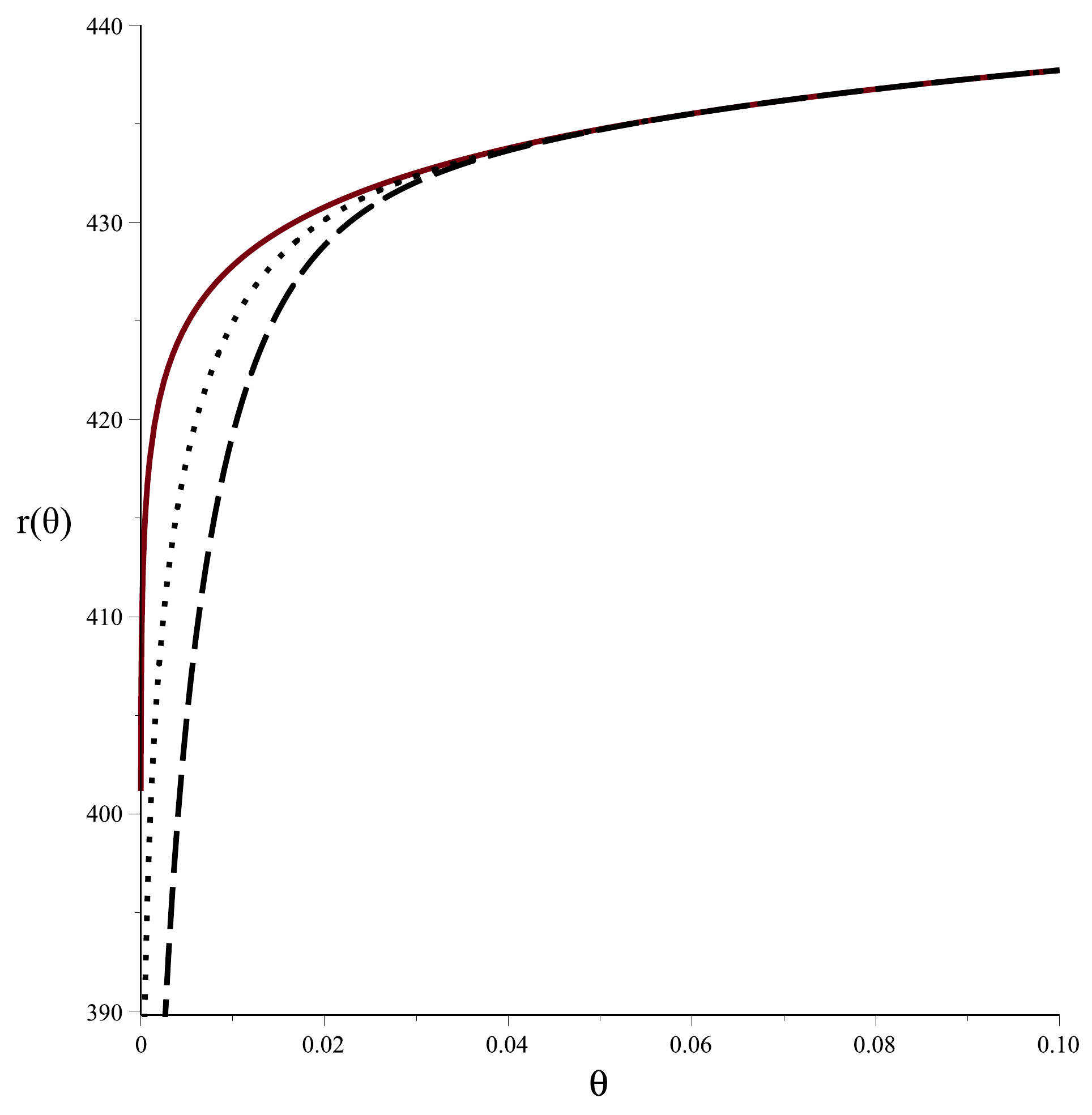}
\caption{Radius $r(\theta)$ as a function of the angle $\theta$. Solid line - matter is absent, dotted line - small density of the scalar field, dushed line - moderate density of the scalar field. Boundary conditions: $r(\pi) = 447, r'(\pi) = 0, R(\pi) =2/r(\pi)^2, R'(\pi) = 0$ and the parameters $R_0 = 10^{-5}; u_1 = 100;$ $m=0.15$.}\label{apple}
\end{figure}

\begin{figure}[t]
\centering
\includegraphics[width=0.7\linewidth]{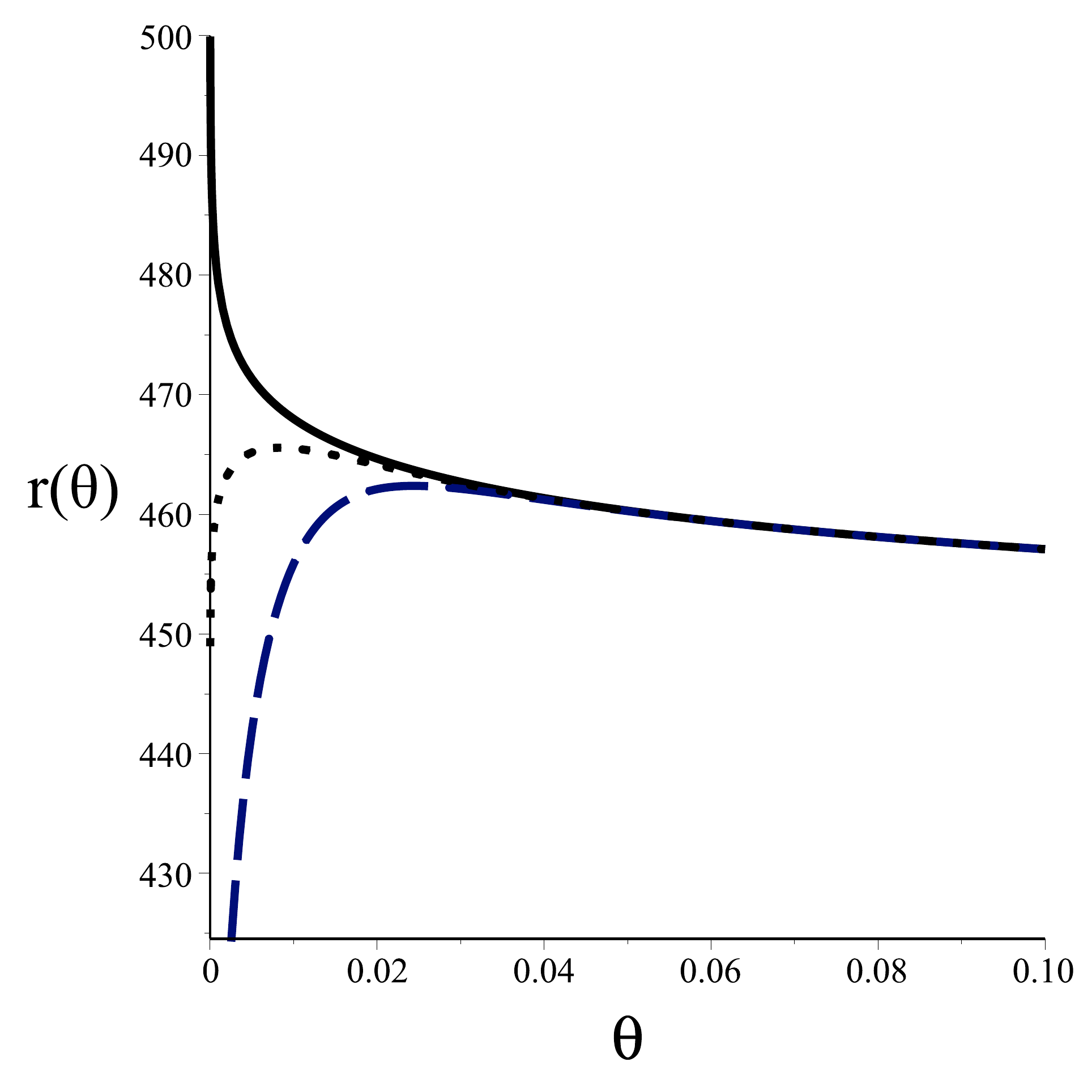}
\caption{Radius $r(\theta)$ as a function of the angle $\theta$. Solid line - matter is absent, dotted line - small density of the scalar field, dushed line - moderate density of the scalar field. Boundary conditions and the parameters the same as in Fig. \ref{apple} except $r(\pi) = 448$.}\label{onion}
\end{figure}

Metric \eqref{metric2} has more general form that causes a thick brane formation - a solutions to Eq.  \eqref{tr-n}.
This equation leads to the trivial equality, if the function $f$ is linear function of $R$. Solutions are much more promising if the function $f$ has more complex form. So suppose that
\begin{equation}\label{fR}
f(R)=u_1 (R-R_0 )^2 .
\end{equation}
As a result explicit form of equation \eqref{tr-n} to be solved numerically is 
\begin{equation}
 \partial^2_{\theta}R+\cot\theta \partial_{\theta} R =- \frac{1}{2}r(\theta)^2 \left(R_0^2 - R^2 \right) -  \frac{r(\theta)^2}{2u_1} T
\end{equation}
Solutions $r_b (\theta)$  for $T=0$ are represented in Figs. \ref{apple} and \ref{onion} (solid lines). Their features are discussed in \cite{Gani}. Due to high nonlinearity of the equation the gravity is able to trap itself in a small region around $\theta =0 $.

It is important to notice that in the framework of ordinary gravity a scale of classical region is about $l_c \geq 1/M_{Pl}$. In our case with D-dim space and multidimensional Planck mass $m_D$ the situation is not so evident. Main difference is found in hyperbolic extra spaces \cite{RuRole}. For a space with positive curvature (like in our case), the result coincides in order of magnitude with the expected one - $l_c \geq 1/m_D (=1 \text{ in } m_D \text{ units})$. This means that the classical solutions represented in Fig.\ref{apple} and \ref{onion} are valid at $\theta\sim l_c/r(\theta) \geq  1/r(\theta)\sim  1/400$. So the quantum effects invalidate our classical results only in close vicinity of the point-like defect at $\theta =0$.

\begin{figure}[t]
\centering
\includegraphics[width=0.7\linewidth]{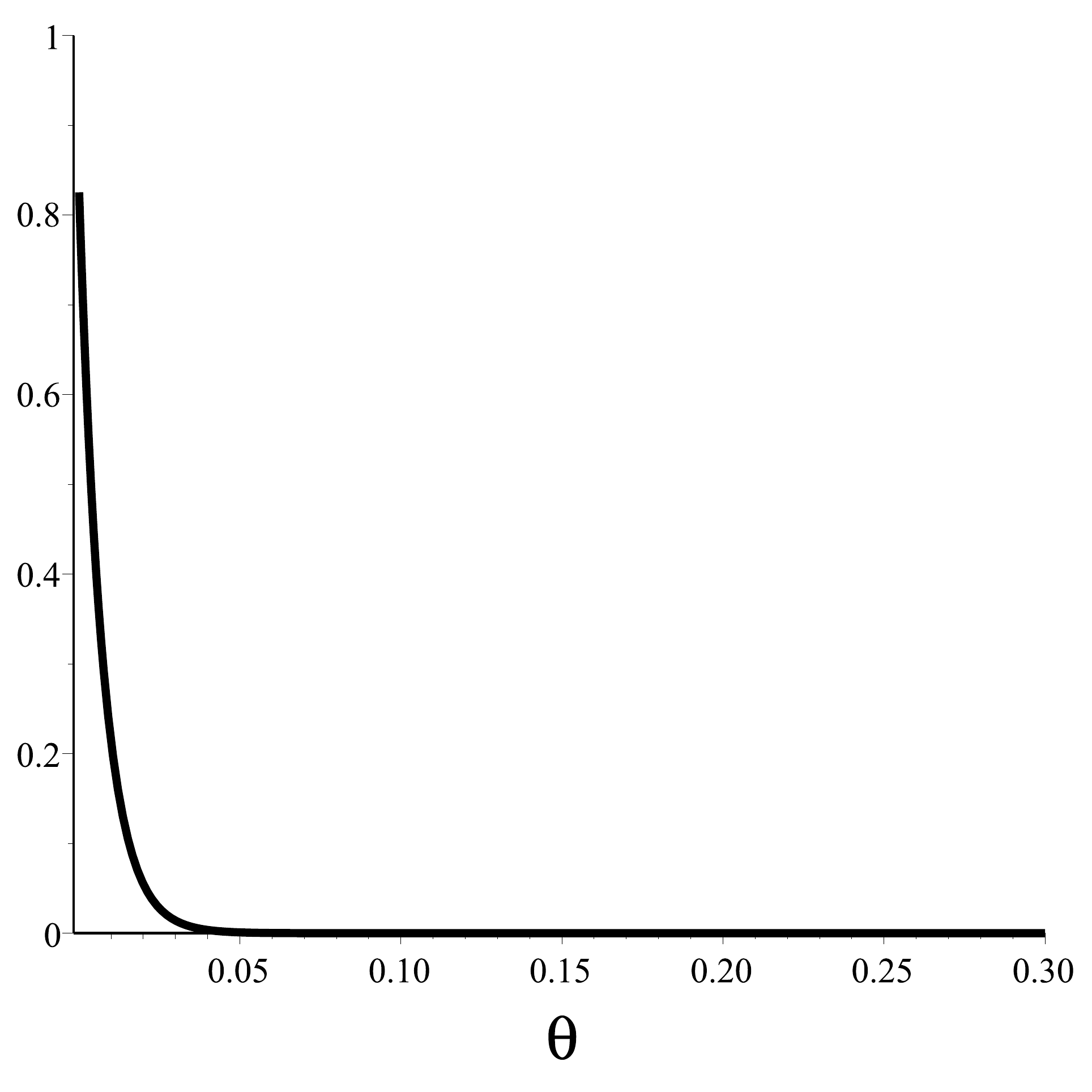}
\caption{Typical distribution of the energy-momentum tensor trace of the scalar field}\label{Mattens}
\end{figure}

The 2-dim metric depends on internal coordinates only. So that it may be interpreted as a thick 4-dim brane embedded into 6-dim space $M_4 \times M_2$. In the framework of the Einstein-Hilbert gravity a brane must include a matter to avoid singularities \cite{Vilenkin}. In our case of a matter unavailability the brane has a predictable structure in classical region and fails near brane. The latter should be true not only for the model in question.

Let us discuss the mechanism of matter concentration inside a small volume of  compact manifold with "apple" - type metric represented in Fig. \ref{apple}.

\section[Loc]{Localization of scalar field}

\subsection{Field trapping by point like defects of 2-dim metric}

Consider the influence of deformed extra geometry represented by solid line in Fig.\ref{apple} on a scalar field distribution.
As it is shown in this section a scalar field with the Lagrangian
\begin{equation}\label{Lscalar}
L_m =\frac12 \partial_a \varphi G^{ab}\partial_b \varphi - \frac{m^2}{2}\varphi^2
\end{equation}
is localized in the vicinity of such defect.

The field is assumed to be uniformly distributed in our 4-dim space,
\begin{equation}
\varphi(x,y) = Y(y) 
\end{equation}
so that the classical equation of motion is
 \begin{equation}\label{eqphi}
\square_n Y(y) +m^2 Y(y) =0.
\end{equation}
With metric \eqref{metric2} it may be written in the following form
\begin{equation}\label{eq3}
\cot(\theta)\partial_{\theta}Y(\theta) + \partial^2 _{\theta}Y(\theta)  - m^2 r_b (\theta)^2Y(\theta) =0.
\end{equation} 
This equation can be simplified to obtain analytical solution. 
For that purpose remind that
characteristic size $\bar{r}$ of an extra space must satisfy the condition
\begin{equation}\label{rmD}
\bar{r}\gg 1/m_D
\end{equation}
which is necessary condition for the extra space to be considered classically. For the particular case represented in Figs.\ref{apple} and \ref{onion}, $\bar{r}\sim \bar{r_b}\simeq 400/m_D$ and inequality \eqref{rmD} holds. 
Let us define  "potential"\, $v(\theta)\equiv m^2 r_b(\theta)^2  \epsilon^2$ where small parameter  $$\epsilon =1/(\bar{r}m_D)\ll 1$$ was introduced.

In the WKB spirit a solution to \eqref{eq3} may be found in the form
\begin{equation}\label{WKB}
Y(\theta) = Ce^{S/\epsilon}
\end{equation}
Here $C$ is a normalization constant. 
The equation acquires the form
\begin{equation}
\epsilon \cot(\theta)S'_{\theta} + \epsilon S'' + S'{}^2_\theta -v(\theta)=0
\end{equation}
The second term may be omitted as compared to $S'{}^2_\theta$. 

Let us demonstrate that the first term is also small in the classical region. Necessary condition for the classical description of an area of a size $l$ reads $l\gg 1/m_D$. In the angular units it means that $\theta \simeq l/r_b (\theta)>l/ \bar{r}_b\gg 1/m_D\bar{r}_b$ ($\theta \ll 1$ is supposed). First inequality is true in close vicinity to $\theta = 0$ where the condition $r_b(\theta)<\bar{r}_b$ takes place as it can be seen from Fig.\ref{apple}. Hence the classical description is valid if $\theta$ satisfies the inequality 
\begin{equation}
\theta\gg \frac{1}{\bar{r}m_D}=\epsilon.
\end{equation} 
This condition is equivalent to the condition $\epsilon \cot (\theta) \ll 1$ at small angles. It means that the first term is also small in comparison with the term $S'{}^2_\theta$ and classical solution to \eqref{WKB} has the following form
\begin{equation}\label{cut}
Y(\theta)= C\exp \{-m\int^{\theta} _0 d\theta' r_b (\theta')\}
\end{equation}
If the radius $r_b (\theta)$ is growing monotonically as in Fig.\ref{apple} (solid line) the scalar field density has sharp peak as it is seen from Fig.\ref{Mattens}.

As the result the matter concentrates near point-like defects of extra space metric. At the same time it is uniformly distributed throughout our 4-dim space $M_4$. Effects of fermions trapping on apple-like brane  is discussed in \cite{Gogber} on the basis of the Einstein-Hilbert gravity.

\subsection{Back reaction of the scalar field to the metric}

Let us study the influence of a scalar field on the metric of 2-dim manifold. Suppose that the trace of energy-momentum tensor has a sharp peak near $\theta =0$ as in Fig.\ref{Mattens}. An origin of such peak does not matter. It could be a result of metric influence as was discussed above or an accidental fluctuation.

The energy-momentum tensor
\begin{equation}\label{Tab}
T_{ab}(\varphi(\theta)))=\partial_a \varphi \partial_b \varphi -G_{ab}\left[\frac{1}{2}\partial_c \varphi G^{cd}\partial_d\varphi -U(\varphi)\right]
\end{equation}
should reveal a sharp maximum as well.

For 2-dim extra space we have
\begin{eqnarray}\label{T}
&& T=G^{ab}T_{ab}=2U(\varphi)= m^2 \varphi^2 = \nonumber \\
&& =C^2 m^2 Y^2 = C^2 m^2 \exp \{-2m\int^{\theta} _0 d\theta' r_b (\theta')\} .
\end{eqnarray}
Numerical calculations with the help of the same equation \eqref{tr-n} and the trace in the r.h.s of Eq. \eqref{T} are represented in Fig.\ref{apple}. It can be seen that local matter distribution strongly influences 2-dim metric
making the wall deeper (dotted and dashed lines). It can be concluded from Fig.\ref{onion} that a matter clamp is able to form a gravitational well. The more densely the matter is concentrated near a point the deeper the gravitational well.
It coincides with our physical intuition. 

\section{Self-fitting of the cosmological $\Lambda$ term and the Planck mass}

The approach developed above is based on approximation \eqref{ll}. Let us check its validity. To be more precise  and keeping in mind the connection $R_4 \sim \Lambda$ we must obtain the smallness of the $\Lambda$ term, $\Lambda \ll R_n,$ see \eqref{density}.

\begin{figure}[t]
\centering
\includegraphics[width=0.7\linewidth]{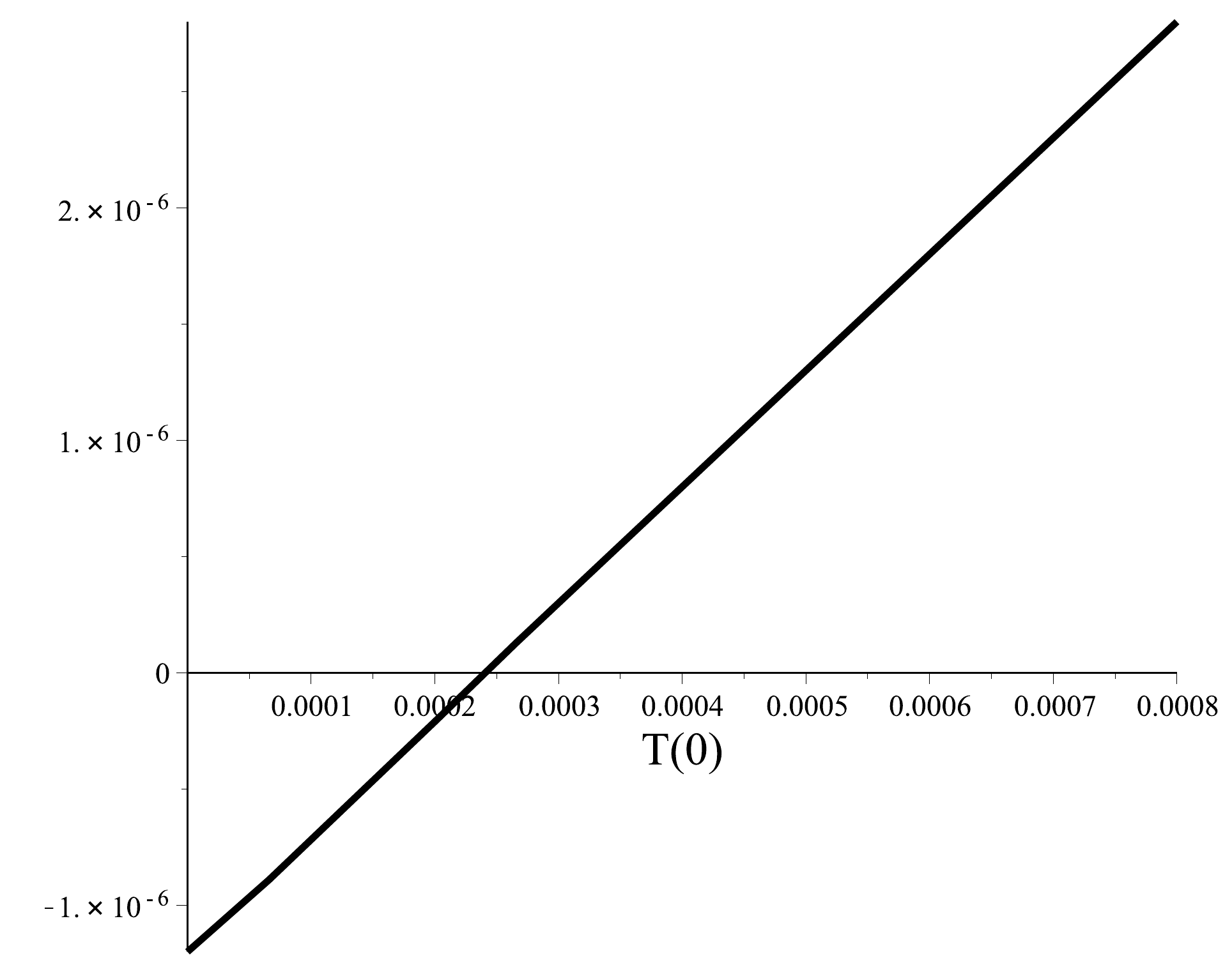}
\caption{The cosmological $\Lambda$ term versus scalar density distributed on the extra space. $T(0)$ is the trace of the energy momentum tensor at $\theta =0$, $m_D=1$}\label{LambdaZero}
\end{figure}

Main assumption of the developed approach is a formation of universes from space-time foam. Any initial conditions may occur with non-zero probability.
In our case it relates to both scalar field amount placed in the extra space and the form of extra space. 
The chosen physical parameter $u_1 >0$ and the quadratic form of the function $f(R)$ means that $f(R)>0$ always. Simple analysis of expression \eqref{Lscalar} indicates that  $L_m <0$ for stationary solutions of the scalar field so that these terms could annihilate each other. The matter distribution strongly depends on initial conditions that can vary  within a wide range. This observation is confirmed by numerical calculations represented in Fig.\ref{LambdaZero}. It is assumed that the trace of energy momentum tensor $T(\theta=0)$ is also the result of initial conditions

Thus, the cosmological constant can vary from negative values (when matter is absent) to positive values (due to matter contribution). The variety of universes differs due to their initial conditions. In particular there exists a set of universes with cosmological $\Lambda$ terms being arbitrarily close to zero. According the Fig.\ref{LambdaZero} these universes are formed with a scalar field distribution such that $T(0)\approx 0.00024$. 

\begin{figure}[t]

\centering
\includegraphics[width=0.7\linewidth]{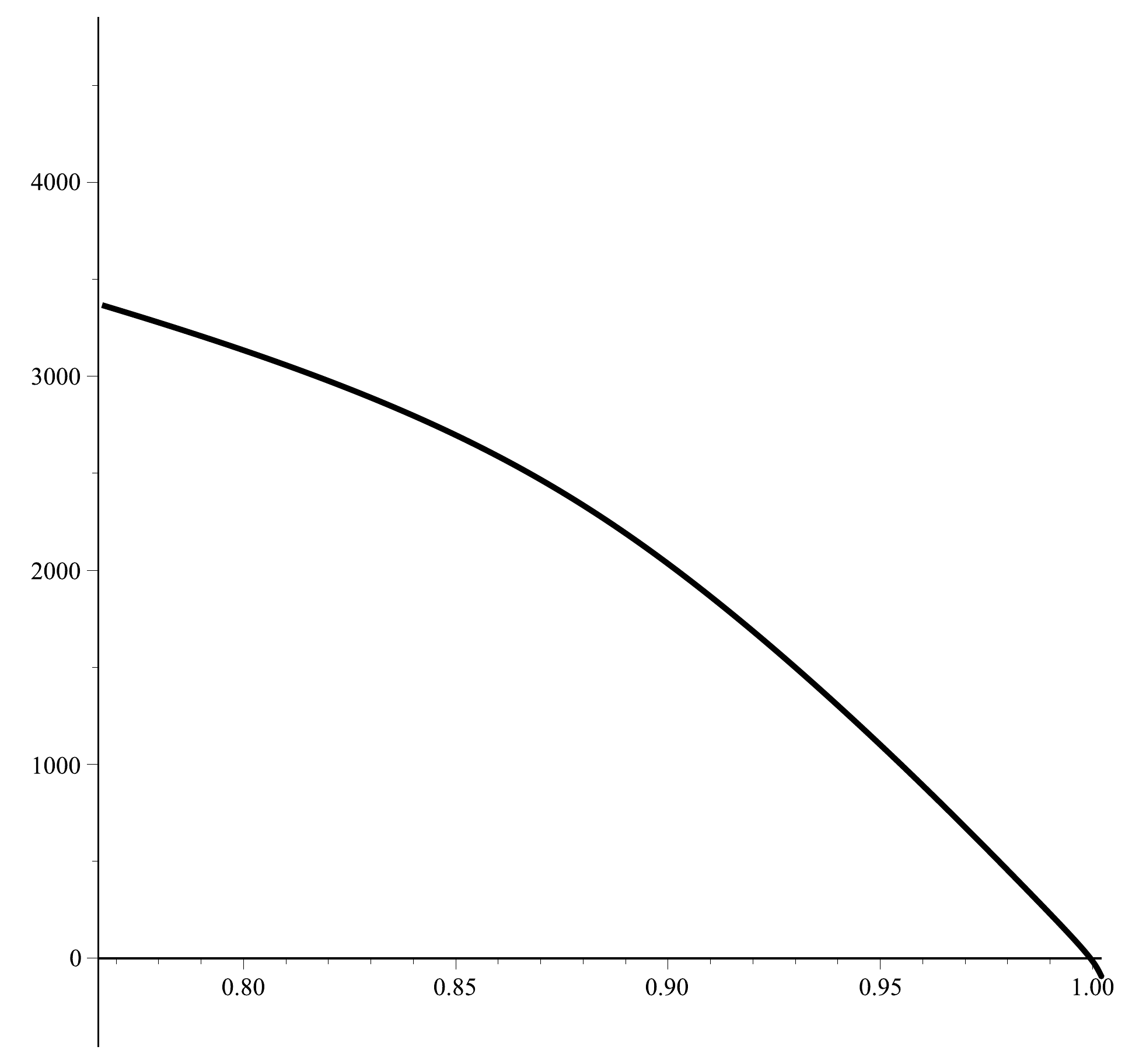}
\caption{The Planck mass square versus boundary value $r(\theta =\pi)$ normalized by the sphere metric  at $r(\pi)=r_0$.}\label{Planck}
\end{figure}

The Planck mass depends significantly on the metric of extra space as shown on Fig.\ref{Planck}. Even if $m_D =const$, the calculated Planck mass can be fitted to the observable one by modification of the boundary conditions. Notice that some boundary conditions lead to negative values of the Planck mass and hence are unacceptable.

\section{Conclusion}

This paper discusses extra spaces with metrics differ from the maximally symmetrical ones within the framework of $f(R)$ gravity. 
A variety of initial conditions caused by the space-time foam leads to a continuous set of 2-dim metrics containing  point-like defects of the extra space. The extra space metric is uniform in our 3-dim space and therefore may be responsible for 4-dim brane formed in 6-dim space $M_6 = M_2 \times M_4$.
The extra space structure is determined everywhere except close vicinity of the brane where the quantum gravitational effects destroy the classical description.

A scalar field distributed in the extra space is localized within the vicinity of the point-like defect. 
Such clump of scalar field, in its turn, makes the gravitational well of the point-like defect more prominent. 

Both the metric and the scalar field contribute to the value of the cosmological $\Lambda$ term. The signs of their contribution are opposite, so they could annihilate each other at some specific initial conditions.  
It is shown that the continuous set of 2-dim metrics contains a subset responsible for the range of $\Lambda$ terms including the zero value. Consequently this subset contains an extra space metric leading to observable value of the cosmological constant.

\section{Acknowledgments}
 The author is grateful to the group of Cosmology NRNU MEPhI  for useful discussions.
This work was supported by the Ministry of Education and Science of
the Russian Federation, Project No. 3035.

\end{document}